\documentclass[12pt]{article}
\usepackage{amsmath}
\usepackage{latexsym}
\usepackage{amsfonts}
\usepackage{amssymb}
\usepackage[dvips]{graphicx}
\usepackage[cp1250]{inputenc}

\title{Note on lattice spin in graphene and "spin from isospin" phenomenon } 
\author { Piotr Kosi\'nski$^{1}$\thanks{supported by  Polish government grant no. 504/092.}, Pawe{\l} Ma\'slanka$^{1}$\thanks{supported by {\L}\'od\'z.University grant no. 506/1037 }, Jadwiga S{\l}awi\'nska$^{1,2}$\thanks{supported by by Polish government (MNiSW) within the contract No. N N202 086040.}, Ilona Zasada$^{2}$\\  
$^1$Department of Theoretical Physics and Computer Science,\\
 University of Lodz, Pomorska 149/153, 90-236 Lodz, Poland,\\
$^2$Solid State Physics Department, University of Lodz,\\
Pomorska 149/153, 90-236 Lodz, Poland }
\date{}

\begin{document}
\maketitle
\begin{abstract}

    It is well-known that the dynamics of low energy electron in graphene honeycomb lattice near the $K/K'$ points can be described, in tight-binding
 approximation, by 2+1 massless Dirac equation. Graphene’s spin equivalent, “pseudospin”, arises from the degeneracy introduced by the honeycomb
 lattice’s two inequivalent atomic sites per unit cell. Mecklenburg and Regan 
( Phys. Rev. Lett. 106 (2011), 116803) have shown that, contrary to 
 the common view, the pseudospin has all attributes of real angular momentum. In some circumstances, the internal symmetries can
 produce an important contribution to angular momentum. 
   This phenomenon has been known for many years in particle physics and called "spin from isospin”. 
    We show that similar mechanism works in the case of lattice pseudospin. 

\end{abstract}

\newpage
Angular momentum is one of the most important notions in both classical and quantum physics. This is due to its intimate connection, 
via Noether theorem, with the isotropy of physical space implying rotational invariance of interaction. The angular momentum 
conservation ( together with other invariance principles, like parity) leads to the selection rules concerning
 emission and absorption processes in atomic physics, particle scattering etc. Mathematics of rotation group $SO(3)\sim SU(2)$\
dictates most of the properties of angular momentum: its spectrum, additional rules, uncertainty relations for components etc.
Another important property of angular momentum is that, in general, it is a sum of orbital and spin parts; the former takes
only integer values while the latter - both integer and half-integer ones. Therefore, once we know that the total angular
 momentum is half-integer it can be taken for granted that the spin part gives nonvanishing contribution. 
 
 The separation of angular momentum into orbital and spin parts cannot be inffered from the structure of $SO(3)$ alone. 
However, it can be demonstrated by referring to the larger symmetry group, Galilean or Poincare one. For example, the structure 
of the unitary representations of the Galilei group (for the Poincare group the situation is slightly more involved) implies
 at once the relevant decomposition of angular momentum. 
 
 There is also another way of introducing the spin contribution to total angular momentum, called colloquially "spin from isospin" 
 \cite{b1} $\div$\ \cite{b3}.
 Some field theories posses rich nonperturbative sector separated from the vacuum by a barier of topological origin. They are
 usually characterized by the existence of large internal symmetry. On the quasiclassical level nonperturbative sectors 
are characterized by the existence of nontrivial solutions to classical equations of motion. These solutions are stabilized
due to the existence of the lower bound for energy expressed in terms of some topological invariants. The semiclassical theory
is obtained by quantizing small oscillations around the classical solution. The symmetry group of the Lagrangian includes
 $SU(2)\times G$, where $G$\ is the group of internal symmetries. However, due to the appearance of nontrivial soliton 
solution the actual symmetry group is smaller. For example, if $G$\ contains $SU(2)$\ it may appear that the soliton solution
is invariant under the $SU(2)$\ diagonal subgroup generated by the sum of rotation generators and internal $SU(2)$\ ones. In 
such a way we obtain additional contribution to angular momentum. As an example consider the gauge theory with isospin group
 as a gauge group. Assume all particles entering the theory are bosons so, superficially, there is no spin one-half 
contribution to angular momentum. Consider the monopole solution as "soliton" one. It is "spherically symmetric" in the sense
 that the rotation can be compensated by the appropriate isospin rotation. So there is isospin contribution to rotation generator.
 Now, if some matter bosons have isospin $\frac{1}{2} $\ they give spin $\frac{1}{2} $ contribution to the angular momentum.
 More specifically \cite{b2}, consider the $SU(2)$\ gauge theory consisting of gauge  field $A^a_\mu $\ and
two matter fields $Q^a$ (isospin one) and $U$ (isospin one half). The model has a magnetic monopole solution of the form
 \begin{eqnarray}
 &&Q^a_{cl}=x^aQ(r) \nonumber \\
&&U_{cl}=0   \nonumber \\
&&A^a_{cl,i}=\epsilon _{iab} x^bW(r) \label{1}
\end{eqnarray}
 the upper indices are the isospin ones while the lower refer to space-time transformation properties. So we are dealing with one 
vector isospin$-\frac{1}{2}$\ field, one spinless  isospin$-1$\ and one spinless isospin$-\frac{1}{2}$\ bosonic fields;
 no fermions enter the Lagrangian.

It is now easy to build the quantum theory of small fluctuations around classical solutions (\ref{1}).
Let $A_q$, $Q_q$\ and $U_q$\ be the quantum fluctuations. To lowest order the fields  $A_q$\ and $Q_q$\ do not mix with $U_q$; 
in the nonrelativistic limit one classifies the states of $U_q$\ by solving the Schr\"odinger equation
 \begin{eqnarray}
 &&HU_q=EU_q    \nonumber \\
&&H=\frac{1}{2m} [p_j-eA^a_{clj}T^a]^2 + binding \; potential \label{2}
\end{eqnarray}

\begin{figure}
\begin{center}
\includegraphics[width=0.72\columnwidth]{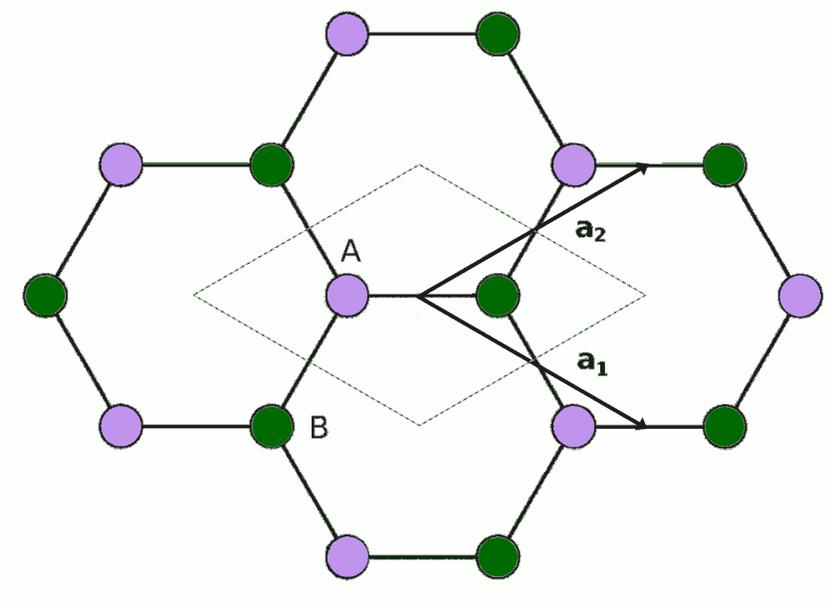}
\end{center}
\caption{\label{siec} Honeycomb crystal lattice of graphene.}
\end{figure} 

It is obvious that $A_{clj}$\ is invariant under simultaneous rotation in ordinary and isospin spaces.
Therefore, the rotational invariance in generalized sense implies that the generator commuting with $H$\ reads
 \begin{eqnarray}
 \vec{J} = \vec{x}\times \vec{p} + \vec{T}   \label{3}
\end{eqnarray}
We see that $\vec{J} $\ gets half-integer contribution through $\vec{T} $ ( $\vec{T} $\ are generators in the isospin 
representation spanned by $U$\ so they corresponds to half-integer representation of $SU(2)$). 
Consider now the electron motion in honeycomb lattice (see Fig.1)

We are interested in nearest-neighbor approximation. It is easy to write out the relevant Hamiltonian \cite{b4}.
Its main ingredient is the single particle Hamiltonian
\begin{eqnarray}
&&\mathcal{H}=-t\left[\begin{array}{cc}
-\frac{\Delta }{t} & 1+e^{-i\vec{Q}\vec{a}_1}+e^{-i\vec{Q}\vec{a}_2}    \\
 1+e^{i\vec{Q}\vec{a}_1}+e^{i\vec{Q}\vec{a}_2} & \frac{\Delta }{t}  \\
 \end{array}\right] ;  \label{4}
\end{eqnarray}
where $t$\ is the hopping parameter while $\Delta $-the energy difference between the sites $"A"$\ and $"B"$.

We are interested in the low-energy limit around the $\vec{K}$\ points of Brillouin zone,
 $\vec{K}^\kappa =\kappa \frac{2\vec{b}_2 +\vec{b}_1}{3} +m\vec{b}_1 +n \vec{b}_2 ,\; \kappa =\pm 1$. 
Expanding around these points we find \cite{b4}: 
\begin{eqnarray}
&&\mathcal{H}_{\kappa} =-t\left[\begin{array}{cc}
\Delta  & \sqrt{3}ta\vec{k} \frac{\kappa \vec{a}_d-i\vec{a}_s}{2}   \\
 \sqrt{3}ta\vec{k} \frac{\kappa \vec{a}_d+i\vec{a}_s}{2} & -\Delta   \\
 \end{array}\right] ;  \label{5}
\end{eqnarray}
where $\vec{k} =\vec{Q} -\vec{K}$.

Let us note the analogy with "spin from isospin" case. As  previously, we are dealing with rotationally invariant theory
with no half-integer angular momentum (if we neglect the processes with spin-flip). The lattice plays here the role of classical 
background  configuration which breaks the rotational invariance (due to the appearance of $\vec{a_s}$\ and $\vec{a_d}$\ in the
Hamiltonian (\ref{5})). However, as in the previous case, the invariance can be quite easily restored. To see this note the identity
\begin{eqnarray}
 \vec{k}_\varphi  \cdot(\kappa \vec{a}_d-i\vec{a}_s)   = \vec{k}\cdot  (\kappa \vec{a}_d-i\vec{a}_s) e^{i\kappa \varphi} \label{6}
 \end{eqnarray}  
where
\begin{eqnarray}
 \vec{k}_\varphi =(k_1 cos\varphi +k_2 sin\varphi ,\;-k_1 sin\varphi +k_2 cos\varphi) \label{7}
 \end{eqnarray}
is the rotated momentum. Now, eqs. (\ref{5}), (\ref{6}) and (\ref{7}) imply
\begin{eqnarray}
 \mathcal{H}(\vec{k})=U^+\mathcal{H}(\vec{k}_\varphi) U(\varphi )   \label{8}
 \end{eqnarray}
where
\begin{eqnarray}
&&U(\varphi )=\left[\begin{array}{cc}
e^{\frac{i\kappa \varphi }{2}}& 0    \\
0 & e^{\frac{i\kappa \varphi }{2}} \\
 \end{array}\right]=e^{i\varphi \kappa \frac{\sigma _3}{2} };  \label{9}
\end{eqnarray}
Note, that $U(\varphi )$\ plays the role of compensating "isospin" transformation. Eq.(\ref{8}) leads to the conservation of 
"total angular momentum" 
 \begin{eqnarray}
 \vec{J} = \vec{x}\times \vec{p} + \kappa \frac{\sigma _3}{2}   \label{10}
\end{eqnarray}
Concluding, we see that the lattice pseudospin emerges because the quantization of electronic motion is performed 
on nontrivial background. There is a close analogy with the "spin from isospin" phenomena where the half-integer spin 
results from quantization on monopole background. It should be also noted that the above picture is valid within 
the low (electron) energy approximation. If higher order terms of the expansion of the Hamiltonian in the powers of $ka$\ 
are taken into account the rotation invariance is broken down to the discrete one (reflecting the geometry of the lattice) 
in a way which prevents its restoration by adding some extra terms (like lattice pseudospin).

\end{document}